\documentclass[12pt]{article}
\pdfoutput=1

\usepackage{graphicx}
\usepackage{bm}
\usepackage{amsmath,amsfonts}
\usepackage{amssymb,xfrac}
\usepackage[bbgreekl]{mathbbol}
\usepackage{cite,color}

\def\be{\begin{equation}}
\def\ee{\end{equation}}
\def\bea{\begin{eqnarray}}
\def\eea{\end{eqnarray}}

\begin{document}

\pagestyle{plain}

\begin{center}
~

\vspace{1cm} {\large \textbf{Worldline as a Spin Chain}}

\vspace{1cm}

Amir H. Fatollahi

\vspace{.5cm}

{\it Department of Physics, Alzahra University, \\ P. O. Box 19938, Tehran 91167, Iran}

\vspace{.3cm}

\texttt{fath@alzahra.ac.ir}

\vskip .8 cm
\end{center}

\begin{abstract}
The general theoretical ground for the models based on the compact angle
coordinates is presented. It is observed that the proper dependence 
on compact coordinates has to be through the group elements and
is achieved most naturally in a discrete-time formulation of the theory.
By the construction, the discrete worldline inlaid by compact coordinates 
resembles the spin chains of magnetic systems. As examples, the models 
based on the groups U(1), $\mathbb{Z}_N$ and SU(2) are explicitly 
constructed and their exact energy spectra are obtained.
As the consequence of minima in the spectra, 
the models exhibit a phase transition of first-order.
The dynamics by U(1) group is attempted to be fitted to the proposed role for monopoles in the dual Meissner effect of confinement mechanism.
\end{abstract}

\vspace{1cm}


\vspace{2cm}
\hfill \texttt{\footnotesize arXiv:1611.08009}

\newpage
\section{Introduction}

It has been known that treating the gauge fields as compact angle variables would reveal 
some non-trivial aspects of gauge field theories  \cite{polya1,polya2,thooft1,lattice}. 
Among studies based on the compactness of gauge fields, the lattice formulation of gauge 
theories has provided an explanation for the confinement mechanism 
as well as a basis for numerical studies at strong coupling regime \cite{lattice,kogut}. 

By now there are specific instances of affinities  
between coordinates and gauge fields. The oldest example 
is the one by the special theory of relativity, by which it is understood that both space-time 
coordinates and gauge fields transform as 4-vectors under the Lorentz transformations. 
As another example, based on the duality proposed in  
\cite{xpsi}, one can formulate the coordinate/field correspondence in both Abelian and
non-Abelian gauge theories \cite{cfdual}.
The other instance of relation between coordinates and gauge fields 
is provided by T-duality of string theory. Accordingly, 
the transverse coordinates of Dp-branes in the dual theory
are represented by the gauge fields of open string states, 
leading to the correspondence \cite{tasi,9510017}
\begin{align}\label{1}
A_i \longleftrightarrow X_i/l_s^2, 
\end{align}
in which $l_s$ is the string theory length. 
At weak coupling the dynamics of $X_i$'s is captured by the
theory resulted from dimensional reduction of the ordinary U(1) gauge theory 
\cite{tasi,9510017}. In particular the reduction on all spatial components of the 
gauge field yields the D0-brane dynamics, namely \cite{tasi}
\begin{align}\label{2}
S_\mathrm{D0}=\int \! dt~\frac{m_0}{2}\, \dot{x}_i^2,
\end{align}
in which $m_0\propto 1/g^2 $ ($g$: gauge coupling) \cite{tasi}. 
The transverse coordinates of $N$ number of Dp-branes are 
represented by $N$ dimensional hermitian matrices \cite{9510135}.

In \cite{0bfath} the dimensional reduction of pure U(1) lattice gauge theory
is considered to model the dynamics of 0-branes at strong coupling regime. 
The model by \cite{0bfath} 
might be considered as the result of combination of 
two themes mentioned earlier, 1) treating gauge fields as compact angle variables \cite{polya1,polya2,thooft1,lattice}, 
2) assuming similar characters between coordinates and gauge fields 
\cite{xpsi,cfdual,tasi,9510017}.
The explicit form of the action after the dimensional reduction of 
U(1) lattice gauge theory is:
\begin{align}\label{3}
S_0=\frac{1}{g^2}\sum_{n}\left(\cos\frac{x_{n+1}-x_{n}}{R}-1\right)
\end{align}
in which the coordinates appear as the compact angle variables depending on 
discrete imaginary time label $n$. 
By this form the worldline theory takes the form of 
the 1D plane-rotator model of spin lattice systems \cite{mattis}.
Based on the prescription for the original lattice gauge theory 
\cite{lattice}, using the transfer-matrix method the quantization of the model is 
formulated. The exact energy spectrum as the function of gauge coupling 
is obtained, with a minimum at critical coupling $g_c=1.125$ in the 
lowest energy. As the direct consequence of the minimum, 
the model exhibits a first-order phase transition between coexistent phases
with small and large couplings \cite{0bfath}. Based on discontinuous nature of the first-order phase transition, for $g< g_c$ and $T\approx 0$ the effective zero mean-square velocity $\langle v^2 \rangle $ is zero. 

The purpose of the present work is to provide a general theoretical 
ground for the models based on compact angle coordinates.
It is clarified that the dependence on the group elements 
rather than the algebra ones would lead to invariance of 
the action under the total shifts in the compact domain, like that
is happening in (\ref{3}).
It is observed that the proper dependence is achieved most naturally in 
a discrete-time formulation of the theory. 
The worldline action by the formulation
resembles the 1d spin chain Hamiltonian of magnetic systems, 
with coordinates appearing as the spin degrees of freedom.  
The transfer-matrix method is used to define the quantum theory \cite{wipf}. 
As here the 1d spin chain is used for a ``particle-like" dynamics
interpretation, there should be a square-root of the mass in the definition
of the transfer-matrix elements \cite{wipf}. As a direct consequence of the presence
of the square-root, opposite to the 1d chains of magnetic systems, here 
the energy spectra develops minima.
In a path-integral representation of the formulation, it is emphasized that 
the square-root pre-factor in the definition of the transfer-matrix elements, 
in contrast to the case with infinite extent coordinates, can not 
be absorbed by a change of the integration variables. 

It is well known that the 1d spin systems with short range 
interactions do not exhibit the second-order phase transition 
expected for these systems. However the present model, as 
the consequence of the minima in the spectrum, exhibits a first-order phase transition. 
As mentioned earlier, although the worldline by the model looks like a 1d spin chain, 
due to the square-root pre-factor in matrix elements the spectra is different.
The phase transition nature by the model will be discussed based on 
the behavior of the Gibbs free energy. In particular, the plot of $G$ versus
the thermodynamical variable $M$, as the effective mean-square-velocity,
develops cusp below a critical temperature $T_c$. 
At the cusp the derivative $\frac{\partial G}{\partial M}$ 
is discontinuous, as expected in a first-order phase transition. 

As examples for the formulation, the models based on the U(1), 
$\mathbb{Z}_N$ and SU(2) groups are explicitly constructed.
In all of the examples the exact energy 
eigen-values are obtained, leading to the first-order phase transition. 

The organization of the rest of the paper is as follows. In Sec.~2 the 
basic assumptions and ingredients for the formulation based on compact
coordinates are presented.
In Secs. 3, 4 and 5 three examples based on the groups
U(1), $\mathbb{Z}_N$ and SU(2) are presented explicitly. For 
all the three groups the energy eigen-values are obtained exactly, 
together with the discussion on the nature of the phase transitions in them.
Sec.~6 is devoted to conclusion and discussion.

\section{Basics and Formulation}

The transition amplitude between positions $x_0$ and $x_{N}$
at times $t_0$ and $t_N$ is represented by the path-integral \cite{sakurai}
\begin{align}\label{4}
\langle x_N,t_N|x_0,t_0\rangle=\lim_{N\to \infty} \int_{-\infty}^{\infty} \prod_{n=1}^{N-1} 
\sqrt{\frac{m}{2\pi \,\mathrm{i}\,\hbar\, \epsilon}}dx_n ~ e^{\mathrm{i}\, S[t_0,t_N]/\hbar}
\end{align}
in which $\epsilon =(t_N-t_0)/N$, tending to zero 
in the limit $N\to \infty$. It is noticed that in above all the intermediate positions
$x_n$'s have infinite extents, $-\infty < x_n <\infty$. 
From the considerations mentioned earlier in Introduction, 
here the main concern is about the finite-extent coordinates. 
The finite-extent coordinates 
are quite known in physics. The most familiar ones are the angle variables of polar 
coordinates in 2D and 3D problems. The other example happens when the system 
is defined inside a finite volume, like a box or a sphere. 
As a definite example, let us consider a free particle on
a circle of radius $R$. Defining the angle variable $\phi=x/R$
with $-\pi \leq \phi \leq \pi$,  we have
\begin{align}\label{5}
L=\frac{1}{2}mR^2\dot{\phi}^2 ~~\to~~ H=\frac{p_\phi^2}{2mR^2}
\end{align}
leading to the eigen-functions and eigen-values as follow
\begin{align}\label{6}
\psi_n(\phi)=\frac{1}{\sqrt{2\pi}} e^{\mathrm{i}\, n \phi},~~~~E_n=\frac{n^2\hbar^2}{2mR^2}
\end{align}
Then the transition amplitude between two different positions is known to be
\begin{align}
\langle \phi_N,t_N|\phi_0,t_0\rangle&=\sum_{n=-\infty}^\infty 
\psi_n(\phi_N)\psi_n^*(\phi_0)\, e^{-\mathrm{i}\,  E_n (t_N-t_0)/\hbar}
\cr
\label{7}
&=\frac{1}{2\pi}\sum_{n=-\infty}^\infty 
e^{\mathrm{i}\,n(\phi_N-\phi_0)} e^{-\mathrm{i}\,n^2 \hbar(t_N-t_0)/2mR^2}
\end{align}
The above can be expressed in terms of the 3rd Jacobi theta function
\begin{align}\label{8}
\vartheta_3(z,\tau )=\sum_{n=-\infty}^\infty e^{\mathrm{i}\, \pi \,\tau \, n^2+2\,
\mathrm{i}\, n\, z}
\end{align}
by which we have \cite{tome}
\begin{align}\label{9}
\langle \phi_N,t_N|\phi_0,t_0\rangle=\frac{1}{2\pi}\vartheta_3
\left(\frac{\phi_N-\phi_0}{2},\frac{-\hbar (t_N-t_0)}{2\pi mR^2}\right)
\end{align}
Using the modular property of $\vartheta_3$
\begin{align}\label{10}
\vartheta_3(z,\tau )=(-\mathrm{i}\, \tau )^{-1/2} e^{-\mathrm{i}\, z^2/\pi \tau }\,
\vartheta_3\left(\frac{z}{\tau },-\frac{1}{\tau }\right)
\end{align}
the transition amplitude recasts as \cite{tome}
\begin{align}\label{11}
\langle \phi_N,t_N|\phi_0,t_0\rangle=\sum_{n=-\infty}^\infty 
\sqrt{\frac{m R^2}{2\pi\,\mathrm{i}\,\hbar (t_N-t_0)}}
\exp\left(\frac{\mathrm{i}}{\hbar} \frac{mR^2 (\phi_N-\phi_0-2\pi n)^2}{2(t_N-t_0)}\right)
\end{align}
in which the summand is easily recognized as the 
transition amplitude of a free particle experiencing the position difference 
$x_N-x_0=R(\phi_N-\phi_0-2\pi n)$ 
during time $t_N-t_0$ \cite{sakurai}.
There is a nice interpretation for the sum as well. As the particle moves on the 
circle from $\phi_0$ to $\phi_N$, it matters how many times 
it rounds the circle. The sum on $n$, the so-called winding number, is responsible 
for taking into account the contributions from different rounds to the 
amplitude. So, nevertheless 
the coordinate $\phi$ has finite extent, practically the particle
may travel long distances $\Delta x = R|\phi_N-\phi_0-2\pi n|$ for 
$n=0,\pm 1,\pm2,\cdots$.
In fact (\ref{11}) may come to a form similar to (\ref{4}) with only 
an extra summation. First let us present a time-sliced form of (\ref{11}) 
\cite{tome}:
\begin{align}
\langle \phi_N,t_N|\phi_0,t_0\rangle=&\lim_{N\to \infty}  ~
\int_{-\pi}^{\pi} \prod_{j=1}^{N-1}
\sqrt{\frac{mR^2}{2\pi \,\mathrm{i}\,\hbar\, \epsilon}}d\phi_j 
\cr
\label{12}
&~ \times  \prod_{l=0}^{N-1} \sum_{n_l=-\infty}^\infty 
\exp\left[\frac{\mathrm{i}}{\hbar}  \frac{mR^2 (\phi_{l+1}-\phi_l-2\pi n_l )^2}
{2\,\epsilon}\right]
\end{align}
It is noticed that at $l$-th time-slice the winding number $n_l$ is introduced \cite{tome}. 
Now, by the following change in the integral variables \cite{tome}:
\begin{align}\label{13}
\sum_{n_l=-\infty}^\infty \int_{(2n_l-1) \pi}^{(2n_l+1)\pi} d\phi_l \to 
\int_{-\infty}^{\infty} d\Phi_l
\end{align}
the expression (\ref{12}) recasts to  \cite{tome}:
\begin{align}\label{14}
\langle \phi_N,t_N|\phi_0,t_0\rangle=&\lim_{N\to \infty} \sum_{n=-\infty}^\infty \int_{-\infty}^{\infty}
 \prod_{j=1}^{N-1} \sqrt{\frac{mR^2}{2\pi \,\mathrm{i}\,\hbar\, \epsilon}}d\Phi_j \cr
&~~~\times \exp\left[\frac{\mathrm{i}}{\hbar} \sum_{i=0}^{N-1} 
\frac{mR^2 (\Phi_{i+1}-\Phi_i)^2}{2\,\epsilon}\right]
\end{align}
in which $\Phi_N=\phi_N$ and $\Phi_0=\phi_0+2\pi n$.
In above all the intermediate angles $\Phi_j$'s 
are integrated over the whole real-line $\mathbb{R}$.
In fact by the change of variable
(\ref{13}), integrating the intermediate angle $\phi_l$ over $[-\pi,\pi]$ together with sum over infinite possible rounds $n_l$ is replaced by integration of $\Phi_l$ over $(-\infty,\infty)$. 

The above example shows that it is insufficient to  
merely use finite-extent coordinates to be granted the non-trivial aspects. 
The key point about the above treatment of motion on circle is that 
the time-sliced form of the action appearing in the path-integral 
(\ref{12}) is not invariant under the multi-round shifts
\begin{align}\label{15}
\phi_l \to \phi_l + 2\pi \, k_l,~~~~k_l=0,\pm1,\pm2,\cdots
\end{align}
with $k_l\neq k_j$ for $l\neq j$. It is noticed that the above shift is directly related to 
the different number of possible rounds on circle represented earlier by $n_l$ in (\ref{12}).
Reminding that the group space of U(1) is circle, we can express 
the main idea in a group theoretical language. Defining the U(1) group
element by $U_\phi=\exp(\mathrm{i}\, \phi)$, the action by the Lagrangian (\ref{5}) can be 
expressed as \cite{tome}
\begin{align}\label{16}
S=\frac{1}{2} mR^2 \int dt \left(U_\phi^\dagger \dot{U}_\phi\right)
\left(U_\phi^\dagger \dot{U}_\phi\right)^\dagger=\frac{1}{2} mR^2\int dt~ \dot{\phi}^2
\end{align}
In above, although the starting point is taken to be the group element
$U_\phi$, the Lagrangian depends in fact on the algebra
element $\phi$. This observation for the group U(1) is general and holds 
for other groups as well \cite{tome}. 
Obviously the situation changes if the action would 
be invariant under the shift (\ref{15}), namely due to  
dependence on group elements $\exp(\mathrm{i}\,\phi_l)$'s 
instead of the algebra elements $\phi_l$'s. Interestingly, once 
the time parameter is assumed to be discrete the desired dependence is obtained.
By taking time as $t_n=n\, a\,$ for some finite value  
$a$ and integer $n$, the worldline looks like a chain or 1D lattice with spacing 
parameter $a$. On the $n$-th site of this chain the angle $\phi_n$ is sitting. 
So at time step $n$ we have $U_n=\exp(\mathrm{i}\,\phi_n)$, by which the 
discrete-time version of action (\ref{16}) is 
\begin{align}
S&=\frac{1}{2}mR^2 \, a^{-1}\sum_n \left(U_n^\dagger (U_{n+1}-U_n)\right)
\left(U_n^\dagger (U_{n+1}-U_n)\right)^\dagger
\cr
&=\frac{1}{2}mR^2 \, a^{-1}\sum_n (U_n^\dagger U_{n+1}-1)
(U_{n+1}^\dagger U_{n}-1)
\cr
\label{17}
&=-\frac{1}{2}mR^2 \, a^{-1}\sum_n (U_n^\dagger U_{n+1}+U_{n+1}^\dagger U_{n}-2)
\end{align}
which obviously keeps invariance under the shift (\ref{15}).
The phenomenon observed here is partially in reverse direction
of what has happened in lattice formulation of gauge theories. Namely, 
once one tries to introduce gauge symmetry to the theory on lattice
the algebra elements $A_\mu$'s are to be replaced by 
the group elements $\exp(\mathrm{i}\, a \,A_\mu)$'s
in the action \cite{lattice}. Here as we were going to keep 
the invariance under the shift (\ref{15}) the natural solution 
appears to be defining the action on discrete-time worldline. 
In the next sections we will use this as a basis for model building.

Before to end this section it is helpful to discuss the prominent role of 
the imaginary time in the quantization of models with compact domain support in the 
sense described in above. In particular let us consider the matrix 
element by (\ref{4}) \cite{sakurai}
\begin{align}\label{18}
\langle x |\widehat{U}| x' \rangle = \langle x | \exp(-\mathrm{i}\,\Delta t\,\widehat{H}/\hbar)\,  | x'
\rangle= \sqrt{\frac{m}{2\pi \,\mathrm{i}\,\hbar\, \Delta t}}\, e^{\mathrm{i}\, S[t,t+\Delta t]/\hbar}
\end{align}
in which $\widehat{U}$ is the unitary time evolution operator, $\widehat{H}$ is the 
Hamiltonian, and $S[t,t+\Delta t]$ is the action between times $t$ and $t+\Delta t$ 
\cite{sakurai}.  The basic observation is that the above representation in terms of the 
action is not possible when the dynamical variables are to take values inside a compact domain 
in the sense mentioned earlier. The reason can be easily understood for a system with
one dynamical variable $\lambda$ (field or coordinate). The 
generalization to systems with more variables is then straightforward. 
The identity $\widehat{U}\widehat{U}^\dagger=\widehat{\mathbb{1}}$
for the unitary evolution operator $\widehat{U}$,
defining $U(\lambda,\lambda'')=\langle \lambda|\widehat{U}|\lambda''\rangle$, 
in the $\lambda$-basis takes the form:
\begin{align}\label{19}
\int_\Lambda d\lambda'' ~ U(\lambda,\lambda'') ~ U^\star(\lambda',\lambda'') = 
\delta(\lambda-\lambda')
\end{align}
in which $\Lambda$ is the compact domain in which $\lambda$ takes values.
Now, by the representation like (\ref{19}), as the integrand consists of only
the ordinary regular functions and not the distribution ones, 
there is no way that the integral over a compact domain 
can develop a $\delta$-function. 
By lacking the representation (\ref{18}) for a unitary time evolution operator,
the alternative is to assume that time is imaginary ($t\to -\mathrm{i}\,t$). 
In the models with discrete time ($\Delta t =a$), by this alternative option 
the one-step unitary operator $\widehat{U}_1=\exp(-\mathrm{i}\,a\widehat{H}/\hbar)$ 
is replaced by the so-called transfer-matrix operator $\widehat{V}$ whose matrix element 
between two adjacent times $n$ and $n+1$ is given by \cite{wipf}
\begin{align}\label{20}
\langle \lambda_{n+1} | \widehat{V} | \lambda_n \rangle \propto 
\sqrt{m}\,\exp\big(S_E(n,n+1)/\hbar\big)
\end{align}
in which $S_E(n,n+1)$ is the Euclidean action. Then by common eigen-states for
$\widehat{H}$ and $\widehat{V}$, the eigen-values of $\widehat{H}$ are defined
by \cite{lattice,wipf}
\begin{align}\label{21}
E_s=-\hbar\, a^{-1}\, \ln v_s
\end{align}
where $v_s$ is the corresponding eigen-value of $\widehat{V}$.
Provided that $\widehat{V}$ does not have negative eigen-values, the
above would give a consistent description of the quantum
theory based on an action with discrete imaginary time \cite{lattice,wipf}.
This approach is exactly what is chosen in lattice formulation of gauge
theories \cite{lattice}, turning space-time to a Euclidean one,
and it will be used in the present work as well.

In the formulation presented in above the discrete worldline is inlaid by 
numbers as spin variables. An interesting extension 
is to consider the case in which the site $n$ on worldline
is equipped with the spin operator $\hat{S}_n$, promoting 
the worldline to a quantum spin chain. As an example,
let us consider the Heisenberg XYZ model, defined by the 
Hamiltonian operator for two arbitrary adjacent sites
\begin{align}
\widehat{H}_\mathrm{XYZ}=-\frac{1}{2} 
\left(\kappa_x\, \hat{S}^x_n\hat{S}^x_{n+1}+\kappa_y\, \hat{S}^y_n\hat{S}^y_{n+1}+
\kappa_z\, \hat{S}^z_n\hat{S}^z_{n+1} \right)
\end{align}
The spin operators in above are not restricted to a specific representation,
and generally belong to the $2s+1$ dimensional representation for
$s=\frac{1}{2},1,\frac{3}{2},\cdots$. 
The transfer-matrix $\widehat{V}$ of the model, in analogy with (\ref{20}), is then defined by  
\begin{align}
\widehat{V}\propto \sqrt{\kappa_x\kappa_y\kappa_z}\, \exp\big(-a\,\hat{H}_\mathrm{XYZ}\big)
\end{align}
The matrix $\widehat{V}$ in above is hermitian by construction, as it should. 
The energy spectrum by the model can be obtained by the prescription (\ref{21}). 
The detailed nature of the spectrum and the phase structure 
by the model based on the quantum spin chain
is not discussed in here, and is left for future studies.

In summary, by the considerations mentioned in above, the followings are 
the basis for model building:
\begin{enumerate}
\item Time is assumed to be discrete and imaginary, taking values $t_n=n\,a$ for integer $n$.
\item The action with discrete time depends on the group elements 
to enhance the features by the compact nature of group. 
\end{enumerate}

\section{U(1) Group}
For the $i$-th direction with coordinate $-\pi R^i \leq x^i \leq \pi R^i$, the U(1) 
group element at $n$-th time step is taken as $U^i_n=\exp(\mathrm{i}\, x^i_n/R)$. Here for simplicity we take all radii $R^i$'s equal to $R$.
Following (\ref{17}) the Euclidean action takes the form
\begin{align}
S_E&=\frac{\kappa}{2} \sum_{n,\, i} \left(U^{i\,\dagger}_{n}\, U^{i}_{n+1} +
U^{i\,\dagger}_{n+1} \, U^{i}_{n}  -2 \right)
\cr
\label{22}
&=\kappa  \sum_{n,i}\left(\cos\frac{x^i_{n+1}-x^i_{n}}{R}-1\right)
\end{align}
The dimensionless constant $\kappa$ is the defining parameter
of the model (we have set $\hbar=c=1$). 
The above, as discussed in previous section, is invariant under the shift:
\begin{align}\label{23}
x^i_n\to x^i_n+2\pi\, k^i_n R,
\end{align}
with $k^i_n$'s as integer numbers. At the first place let us check the limits:
\begin{align}\label{24}
\begin{split}
x/R& \ll 1\cr
x_{n+1}-x_n&\to a\, \dot{x}\cr
\sum_n&\to a^{-1}\int\! dt
\end{split}
\end{align}
leading to
\begin{align}\label{25}
S_E\simeq -\,\frac{a\kappa}{2R^2}\int dt ~\dot{x}^{2}_i
\end{align}
which describes the dynamics of an ordinary free particle with mass
$m_0=a\kappa/R^2$ in the imaginary time formalism.
As mentioned in Introduction, the action (\ref{22}) is used
in \cite{0bfath} to model the dynamics of 0-branes at strong coupling limit. 
The action (\ref{22}) is the result of dimensional reduction of 
U(1) lattice gauge theory along spatial directions, by setting:
\begin{align}\label{26}
\begin{split}
\kappa&=1/g^2\cr
a\, A^i&\to  x^i/R
\end{split}
\end{align}
The expression (\ref{22}) for the action is also known as the 1D plane-rotator 
model of magnetic systems \cite{mattis}, although here it is interpreted as a discrete 
worldline equipped by the angle variable coordinates $x^i$'s. In this section we 
review the construction by \cite{0bfath}. 
It is useful to define the new variables
\begin{align}\label{27}
y^i= x^i/R
\end{align}
taking values in $[-\pi,\pi]$, by which the action (\ref{22}) takes the form
\begin{align}\label{28}
S_0=\kappa \sum_{n,i}\left(\cos(y^i_{n+1}-y^i_{n})-1\right)
\end{align}
As the action is fully separable for each direction, it is sufficient to consider
only one copy, dropping the index $i$ hereafter.
As mentioned in Sec.~2, the action with discrete
imaginary time can be used to define the quantum theory based on
the transfer-matrix $\widehat{V}$, defined by its matrix elements
\begin{align}\label{29}
\langle  y_{n+1} |\widehat{V} |  y_n\rangle= \sqrt{\frac{\kappa}{2\pi}}
\,\exp\left[\kappa \left(\cos( y_{n+1}- y_n)-1\right)\right]
\end{align}
in which, recalling $m_0\propto \kappa$, 
the normalization prefactor has to be inserted to match the propagator 
(\ref{18}) (see also (\ref{20})) \cite{wipf}
\begin{align}\label{30}
\langle x_2,t_2|x_1 ,t_1\rangle \propto \sqrt{\frac{m_0}{2\pi}}\exp
\left(\frac{-m_0(x_2-x_1)^2}{2\,(t_2-t_1)}\right)
\end{align} 
Using the identity for the modified Bessel function of the first kind:
\begin{align}\label{31}
\exp[\kappa\cos( y'- y)]=\sum_{s=-\infty}^\infty I_s(\kappa) \, e^{\mathrm{i}\,
s\,( y'- y)}
\end{align}
we have for (\ref{29})
\begin{align}\label{32}
\langle  y_{n+1} |\widehat{V} |  y_n\rangle= \sum_{s=-\infty}^\infty
 \sqrt{\frac{\kappa}{2\pi}}e^{-\kappa}I_s(\kappa)\, e^{\mathrm{i}\,s\,
( y_{n+1}- y_{n})}
\end{align}
by which one reads the normalized plane-wave
\begin{align}\label{33}
\psi_s(x)=\frac{1}{\sqrt{2\pi}}\exp(\mathrm{i}\,s\, y),~~~~~~~
-\pi\leq y \leq \pi
\end{align}
as eigen-function with the eigen-value
\begin{align}\label{34}
v_s(\kappa)=\sqrt{2\pi\kappa} \, e^{-\kappa}I_s(\kappa) 
\end{align}
By the known properties of $I_s$-functions we have
$v_s=\sqrt{2\pi \kappa}\,e^{-\kappa}I_s(\kappa)\geq 0$.
This guaranties that the transfer-matrix method defined by (\ref{20}) 
and (\ref{21}) would lead to a consistent quantum 
theory. Also by $I_s(z)=I_{-s}(z)$ the spectrum is doubly 
degenerate for $s\neq 0$. The energy eigenvalues are found by (\ref{21})
and (\ref{34}) 
\begin{align}\label{35}
E_s(\kappa)=-\frac{1}{a} \ln \left[\sqrt{2\pi\kappa} \, e^{-\kappa}I_s(\kappa)  \right]
\end{align}
The behavior of above at the limit $\kappa \to\infty$
can be checked by the saddle point approximation for Bessel functions
\begin{align}\label{36}
I_s(\kappa)=\lim_{\kappa\to\infty}\frac{1}{2\pi}\int_{-\pi}^\pi d y~ \exp(\kappa\cos y +\mathrm{i}\,s\, y)
\simeq \frac{e^\kappa}{\sqrt{2\pi \kappa}} \exp\left(-\frac{s^2}{2\kappa}\right)
\end{align}
by which for (\ref{35}) we obtain
\begin{align}\label{37}
E_s\simeq  \frac{s^2}{2a\kappa}
\end{align}
matching the energy $E=p^2/(2m_0)$ of a free particle with 
momentum  $p=s/R$ along the compact direction,
and mass  $m_0=\kappa\, a/R^2$ by (\ref{25}). 
So in the limit
$\kappa\to\infty$ the spectrum approaches to that of an ordinary particle. For the intermediate $\kappa$ the spectrum is discrete.
In the limit $\kappa\to 0$, using
\begin{align}\label{38}
I_s(z)\simeq \frac{1}{s!}\left(\frac{z}{2}\right)^s,~~~~z\ll 1
\end{align}
we have
\begin{align}\label{39}
E_s=-(s+\frac{1}{2})\,\frac{\ln \kappa}{a}+O(s\ln s) +O(\kappa)
\end{align}
in which the 2nd term is independent of $\kappa$. Also at $\kappa\to 0$
\begin{align}\label{40}
E_{s+1}-E_s\simeq -\frac{\ln \kappa}{a} \gg \frac{1}{a}
\end{align}

\begin{figure}[t]
	\begin{center}
		\includegraphics[scale=1]{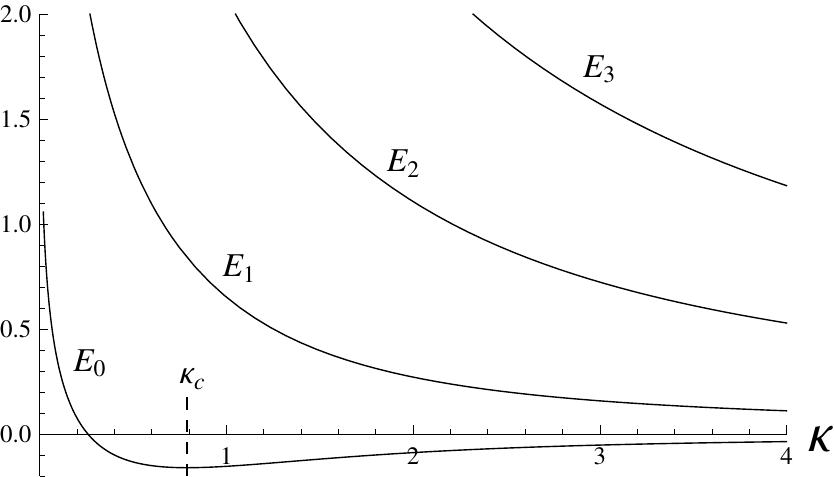}
	\end{center}
	\caption{The few lowest energies by (\ref{35}) versus $\kappa$
		($E$ unit: $a^{-1}$).}
\end{figure}

The interesting observation by the spectrum (\ref{35}) is 
about the energy of ground-state, which has a minimum at $\kappa_c=0.790$; see Fig.~1. As expected the existence of minimum leads to a 
first order phase transition. The one-particle partition function may be evaluated by the definition
\begin{align}\label{41}
Z_1(\beta,\kappa):=\sum_{s=-\infty}^\infty e^{-\beta\, E_s(\kappa)}
\end{align}
or by means of the transfer-matrix operator ($\beta$ in $a$ units) \cite{wipf}
\begin{align}\label{42}
Z_1(\beta,\kappa)=\mathrm{Tr} \,\widehat{V}^\beta=
\int_{-\pi}^\pi \prod_{m=0}^{\beta -1}\sqrt{\frac{\kappa}{2\pi}}\, d y_m
\exp\left[\kappa\sum_{n=0}^{\beta-1} \left(\cos( y_{n+1}- y_n)-1\right)\right]
\end{align}
supplemented by the periodic condition $y_0=y_\beta$.
In the present case the equivalence of (\ref{41}) and (\ref{42}) is 
checked by numerical evaluations. The basic observation by the compact angle variable in above is, in contrast to the situation with infinite extent coordinates, the normalization factor can not be absorbed by a change of integration variable. As the minimum of $E_0$ is in variable $\kappa$, we need the thermodynamical conjugate variable $M$, defined by ($T=\beta^{-1}$)
\begin{align}\label{43}
{M}(\beta,\kappa):= T \,\frac{\partial \ln Z_1(\beta,\kappa)}{\partial\; \kappa}
\end{align}
which is also interpreted as the equation-of-state of the system.
The Gibbs free energy can represent the exact nature of the 
phase transition, 
\begin{align}\label{44}
G_1=A_1+\kappa\,M
\end{align}
in which $A_1=-T\ln Z_1$ is the free energy per particle. 
\begin{figure}[t]
	\begin{center}
		\includegraphics[scale=1]{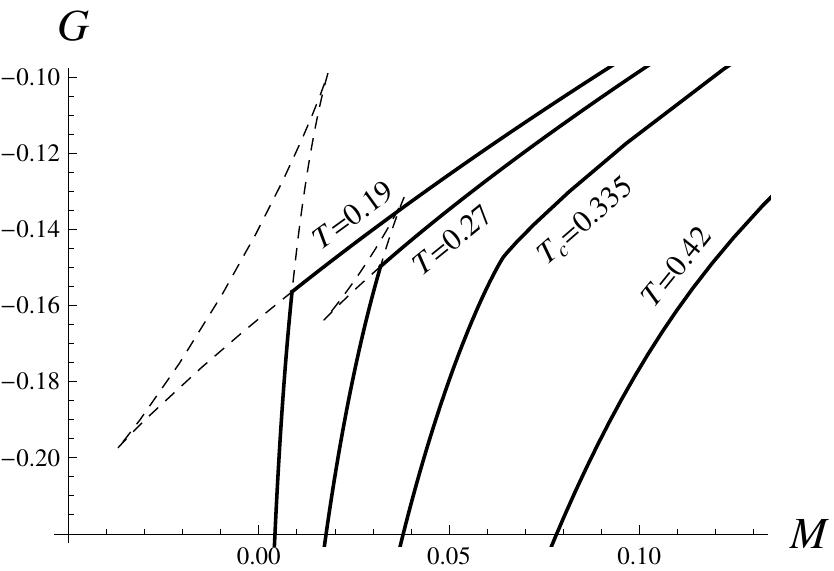}
	\end{center}
	\caption{The $G$-$M$ plots at four temperatures. The dashed pieces are
		not followed by the system due to the minimization of $G$.}
\end{figure}
The isothermal $G$-$M$ plots are presented in Fig.~2. Evidently below the critical temperature $T_c=0.335~a^{-1}$ the plots develop cusps, at which by the minimization of $G$ at equilibrium, the system follows the path with lower $G$ (solid-lines in Fig.~2). As the consequence, for $T<T_c$ there is a jump in first derivative of $\partial G/\partial M$, indicating that the phase transition is a first order one. 
It is evident by now that the above phase structure is quite similar to 
the gas/liquid transition, for which $G$-$P$ plots show exactly 
the same behavior. In a similar way the equation-of-state (\ref{43}) should be modified 
by the so-called Maxwell construction for $P$-$V$ diagram, 
by which during isothermal condensation the pressure (here $M$) is fixed. 
The results of the Maxwell construction for the present model are plotted as
isothermal $M$-$\kappa$ curves in Fig.~3. The flat part at $T_c$ corresponds to 
values:
\begin{align}\label{45}
T_c=0.335:~~\kappa^*=1.403,~~~M^*=0.064.
\end{align}

\begin{figure}[t]
	\begin{center}
		\includegraphics[scale=1.1]{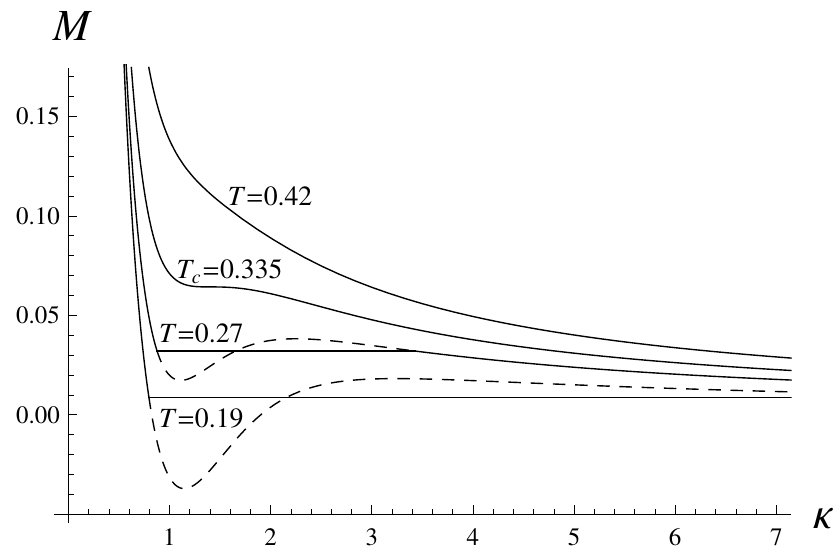}
	\end{center}
	\caption{The isothermal $M$-$\kappa$ plots. The straight-lines are due to the Maxwell construction,
		replacing the dashed parts.}
\end{figure}

\noindent For isothermal curves below $T_c$, the straight horizontal parts describe the coexistent phases of lower and higher $\kappa$'s during the phase transition. 
The interesting fact about the equation-of-state modified by Maxwell construction is that $M$ always remains non-negative, that is $M\geq 0$.  This is specially important by 
expectations from the variable $M$ at the limit $\kappa\gg 1$, at which we expect the ordinary behavior for particles. At this limit, back to
(\ref{25}) and (\ref{37}), we have
\begin{align}\label{46}
M \simeq \frac{1}{2} \langle \dot{y}^2 \rangle \propto \frac{T}{m_0}
\end{align}
where the proportionality is by the properties of free ordinary particles. 
In fact the asymptotic tails in Fig.~3 for large $m_0\propto \kappa$ 
are explained by (\ref{46}). 
There are also asymptotes at $\kappa\to 0$, although with different slopes. In fact the main difference between the case with present model and that of ordinary particles is about the existence of a phase transition. In particular, by the present model and below the critical temperature $T_c$, the two asymptotes by large and small masses (large and small $\kappa$'s)
are connected with a first-order phase transition. 

One may define the order parameter for the present
model as well. For the ordinary magnetic systems with 2nd
order phase transition the order parameter is the
magnetization as the derivative of $G$ \cite{huang}. The 
nonzero magnetization is interpreted as the magnetic ordering phase. However, due to the different nature of the 1st order phase transitions, the derivative of $G$ is discontinuous at cusps in Fig.~2. 
This situation is again quite similar to the case with gas/liquid system, in which the volume difference of coexisting
phases, as the jump in the derivative $\partial G/\partial P$, 
is taken as the order parameter \cite{huang}. 
Similarly, in the present case the jump in $\partial G/\partial M$ defines the 
order parameter, being simply the difference of $\kappa$'s of coexisting phases ($\kappa$-difference at ends of fixed-$M$ line in Fig.~3). 
Like the gas/liquid system, the order parameter tends to zero 
at the critical point (\ref{45}), and larger values of 
order parameter (larger $\kappa$-difference) 
at lower temperatures corresponds to lower fixed-$M$ line in Fig.~3.
In terms of the magnetic ordering $y_n\simeq y_{n+1}$, 
this is the expected behavior with $M\propto \langle \dot{y}^2\rangle \simeq 0$ by (\ref{46}).

\section{$\mathbb{Z}_N$ Group}
The coordinates with a compact domain may form a discrete group such as
$\mathbb{Z}_N$. In this case the worldline looks like a spin chain 
with discrete spin degrees sitting on its sites. In general the worldline 
resemble the spin chain of Potts model ($\mathbb{Z}_2$
as of the Ising model). The members of 
group $\mathbb{Z}_N$ are presented by 
\begin{align}\label{47}
\{1,\varrho,\varrho^2,\cdots,\varrho^{N-1}\}
\end{align}
in which 
\begin{align}\label{48}
\varrho=\exp(\mathrm{i}\,2\pi/N),~~~~~\varrho^N=1
\end{align}
At time-step $n$ the position may be represented by $r_n$ as
\begin{align}\label{49}
U_n=\varrho^{r_n}=\exp(\mathrm{i}\,2\pi\,r_n/N),~~~~~r_n=0,1,2,\cdots,N-1
\end{align}
by which the action takes the form
\begin{align}
S_E=\frac{\kappa}{2}\sum_n \left(
U^{\dagger}_{n}\, U_{n+1} +
U^{\dagger}_{n+1} \, U_{n}  -2 \right) \cr
\label{50}
=\kappa \sum_n \left( \cos \frac{2\pi(r_{n+1}-r_n)}{N} -1 
\right)
\end{align}
The action is invariant under the shifts by $k_n$ being any integer
\begin{align}\label{51}
r_n\to r_n+k_n\, N
\end{align}
It is convenient to define the new variable
\begin{align}\label{52}
\frac{2\pi\,r}{N} = w = 0,\frac{2\pi}{N},\frac{4\pi}{N},\cdots,\frac{(N-1)2\pi}{N}
\end{align}
by which the action comes to the form:
\begin{align}\label{63}
S_E=\kappa \sum_n \left( \cos(w_{n+1}-w_n) -1  \right)
\end{align}
The transfer-matrix element then easily reads
\begin{align}\label{54}
\langle  {n+1} |\widehat{V} |  n\rangle &=\frac{1}{N} \sqrt{2\pi\kappa}
\,\exp\left[\kappa  \left( \cos(w_{n+1}-w_n) -1  \right)\right] 
\\
\label{55}
& =\sum_{s=0}^{N-1} e^{-a\, E_s} ~\frac{1}{N}e^{\mathrm{i}\,s(w_{n+1}- w_{n})}
\end{align}
in which the plane-waves
\begin{align}\label{56}
\psi_s(w)=\frac{1}{\sqrt{N}}\, e^{\mathrm{i}\,s\,w}
\end{align}
satisfy the ortho-normality condition
\begin{align}\label{57}
\sum_w \psi_s^\star (w) \psi_{s'}(w) = \delta_{ss'}
\end{align}
\begin{figure}[t]
	\begin{center}
		\includegraphics[scale=0.9]{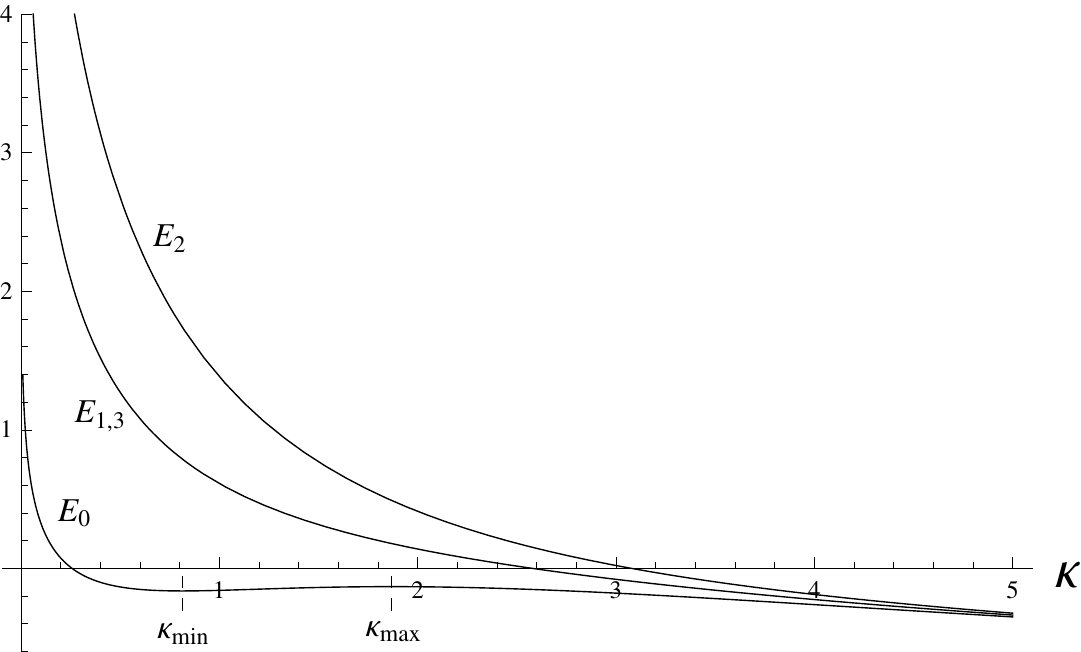}
	\end{center}
	\caption{The energies by (\ref{59}) for $\mathbb{Z}_4$ group.  }
\end{figure}
By the identity (\ref{31}) for Bessel functions, and using the condition (\ref{57}),
the sum on Bessel functions can be partitioned into $N$ cyclic ones, leading to
\begin{align}\label{58}
e^{-a\, E_s}=\sqrt{2\pi\kappa} \, e^{-\kappa} 
\sum_{q=-\infty}^\infty I_{s+qN}(\kappa)
\end{align}
in which the sum is converging due to properties by $I_s$-functions.
So the energy eigen-values are simply given by
\begin{align}\label{59}
E_s(\kappa)=-a^{-1} \,\ln \left[ \sqrt{2\pi\kappa} ~ e^{-\kappa} 
\sum_{q=-\infty}^\infty I_{s+qN}(\kappa)
\right]
\end{align}
for $s=0,1,2,\cdots, N-1$. Due to $I_s=I_{-s}$ we have the following degeneracy:
\begin{align}\label{60}
E_s=E_{N-s}
\end{align}
In fact the lowest and highest eigen-values are as follow
\begin{align}\label{61}
\begin{split}
E_\mathrm{min}&=E_0,\\
E_\mathrm{max}&=E_{N/2},~~~~~~~~N:\mathrm{even} \\
E_\mathrm{max}&=E_{(N\pm 1)/2},~~~N:\mathrm{odd}
\end{split}
\end{align}
\begin{figure}[t]
	\begin{center}
		\includegraphics[scale=1.1]{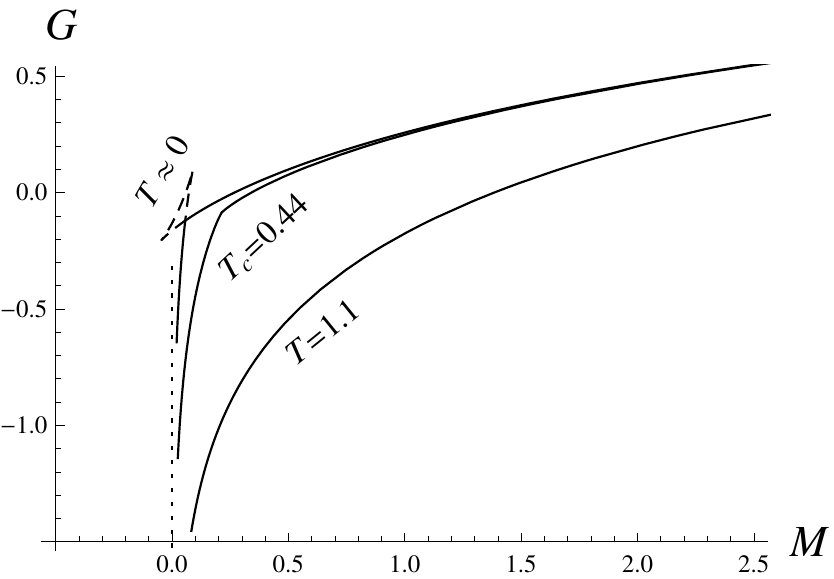}
	\end{center}
	\caption{The $G$-$M$ plots for $\mathbb{Z}_4$ group.  }
\end{figure}
In the limit $N\to\infty$ we expect to recover the 
spectrum (\ref{35}) by U(1) group. It is in fact the case 
using 
\begin{align}\label{62}
\frac{I_{\pm\infty}}{I_\mathrm{finite}} \to 0
\end{align}
It can easily be checked that for $N=2~ \&~ 3$ there is no extrema in $E_0$. For
$N\geq 4$ there are both minimum and maximum. 
The energy eigen-values are plotted in Fig.~4 for the group $\mathbb{Z}_4$, in which 
the extrema are at 
\begin{align}\label{63}
\kappa_\mathrm{min}=0.815,~~~~~~~~\kappa_\mathrm{max}=1.87
\end{align}
Again the one-particle partition function (\ref{41}), together with the thermodynamical functions 
(\ref{43}) and (\ref{44}), can be defined for the system.
Like the case with U(1) group, the appearance of minimum leads to a 
phase transition. The isothermal $G$-$M$ plots for the $\mathbb{Z}_4$ are plotted in Fig.~5.
As expected below the critical temperature $T_c=0.44$ the plots develop cusps, exhibiting 
a first order phase between two co-existing phases with low and high $\kappa$'s. 
However, at $T\approx 0$ there is a difference between the $\mathbb{Z}_N$ and
U(1) groups, that is here there is a finite higher $\kappa$ at which 
the systems follows the path with finite $M$'s. This behavior is 
evident by $T\approx 0$ curve in Fig.~5, in contrast to 
U(1) group with $M=0$ toward $\kappa\to\infty$ at $T\approx 0$.
The two critical $\kappa$'s of $T\approx 0$ for $\mathbb{Z}_4$ group 
at which the cusp starts and ends are as follow 
\begin{align}\label{64}
\kappa_{c1}=0.69,~~~~~~~\kappa_{c2}=8.8
\end{align}

\section{SU(2) Group}
As the case for a non-Abelian group here we consider the SU(2) group in 
one spatial direction. Then at time-step $n$ the group element is 
represented by 
\begin{align}\label{65}
U_n=\exp(\mathrm{i}\,\bm{x}_n\cdot \bm{\sigma}/2R)
\end{align}
in which $\bm{x}_n=(x^1_{n},x^2_{n},x^3_{n})$ represents the three components in the SU(2) sector, and 
$\bm{\sigma}=(\sigma_1,\sigma_2,\sigma_3)$ as Pauli matrices. The action then simply comes to the form 
\begin{align}\label{66}
S_E=\frac{\kappa}{2} \sum_{n} \mathrm{Tr} \left(U^\dagger_{n}\, U_{n+1} +
U^\dagger_{n+1} \, U_{n} \, -\,  2\, \mathbb{1}_2 \right)
\end{align}
in which $\mathrm{Tr}$ is trace over the matrix structure, with $\mathrm{Tr}(\sigma_\alpha\sigma_\beta)=2\,\delta_{\alpha\beta}$.
As the requirement mentioned earlier, the above action is invariant under the shift:
\begin{align}\label{67}
|\bm{x}_n|\to |\bm{x}_n|+4\pi k_n R,
\end{align}
with $k_n$'s as integer numbers. Using the identity
\begin{align}
\frac{1}{2}\,\mathrm{Tr}\,(e^{\mathrm{i}\,\bm{x}_{n+1}\cdot \bm{\sigma}/2R}\,
e^{-\mathrm{i}\,\bm{x}_n\cdot \bm{\sigma}/2R})&=\cos\frac{r_{n+1}}{2R}\cos\frac{r_n}{2R}+\,\bm{\hat{x}}_{n+1}\cdot\bm{\hat{x}}_n\,
\sin\frac{r_{n+1}}{2R}\sin\frac{r_n}{2R}
\cr
\label{68}
&=: \cos\frac{\gamma_{n+1,n}}{2}
\end{align}
in which $r_n=|\bm{x}_n|$ and $\bm{\hat{x}}_n=\bm{x}_n/r_n$, the actions is simplified as 
\begin{align}\label{69}
S_E=2 \kappa \sum_{n} \left( 
\cos\frac{\gamma_{n+1,n}}{2} -1 \right)
\end{align}
In the limits $r_n\, \& \, r_{n+1} \ll R$ we have
\begin{align}\label{70}
\gamma_{n+1,n}^2\simeq \frac{1}{R^2}(\bm{x}_{n+1}-\bm{x}_n)^2
+O\left(\frac{r}{R}\right)^4
\end{align}
by which we have in the continuum limit 
\begin{align}\label{71}
S_E\simeq -\frac{a\kappa}{4R^2} \int dt~\dot{\bm{x}}^2
\end{align}
In above again the minus sign is due to use of imaginary time in the formalism.
The action (\ref{71}) represents the free motion of
a free particle with mass $m_0=a\kappa/(2R^2)$. 

As before, the action (\ref{69}) can be used to define the quantum theory 
based on the transfer-matrix method. 
The group manifold of SU(2) is known to be the 
3-sphere $S^3$, for which the initial parametrization $\bm{x}=(x_1,x_2,x_3)$
in the spherical coordinates $\bm{x}=(r,\theta,\phi)$ is defined 
in the intervals 
\begin{align}\label{72}
0 \leq r \leq 2\pi R,~~~~~~~0\leq \theta \leq \pi,~~~~~~ 
0\leq\phi\leq 2\pi
\end{align}
Also it is convenient to use the replacement
\begin{align}\label{73}
\chi=\frac{r}{2R},~~~~~~~~~ 0\leq \chi \leq \pi
\end{align}
By the above parametrization the SU(2) invariant measure takes the form \cite{su2haar}
\begin{align}\label{74}
d\Omega_3=\sin^2 \!\chi \, d\chi\, d\Omega,~~~\mathrm{with}
~~~~d\Omega=\sin\theta \, 
d\theta\, d\phi
\end{align}
satisfying 
\begin{align}\label{75}
\int_{S^3} d\Omega_3 = 2 \pi^2 
\end{align}
As the case with U(1) group, by the complete orthonormal spherical harmonics on 
$S^3$ as the eigen-functions, we can read the eigen-values of 
the matrix $\widehat{V}$. The normalized spherical harmonics 
on $S^3$ are known to be \cite{su2haar}
\begin{align}\label{76}
\mathbb{Y}_{s \ell m}(\chi,\Omega)=\sqrt{\frac{2^{2\ell +1}(s+1)\, (s -\ell )!\, \ell !^{\,2}}{\pi (s+\ell +1)!}}\sin^\ell  \!\chi~C^{(\ell +1)}_{s-\ell }\!(\cos\chi)~Y_{\ell m}(\Omega)
\end{align}
in which $C^{(\ell +1)}_{s-\ell }(x)$ are the Gegenbauer polynomials and 
$Y_{\ell m}(\Omega)=Y_{\ell m}(\theta,\phi)$ are the ordinary spherical harmonics
on 2-sphere $S^2$. In above all indices are integers and obey the ordering \cite{su2haar}
\begin{align}\label{77}
|m|\leq \ell  \leq  s=0,1,2,\cdots
\end{align}
The above harmonics are the normalized ones \cite{su2haar}
\begin{align}\label{78}
\int_{S^3}  d\Omega_3~ \mathbb{Y}_{s\ell m}(\chi,\Omega)~ 
\mathbb{Y}_{s'\ell' m'}^\star(\chi,\Omega)=
\delta_{ss'}\,\delta_{\ell \ell '}\,\delta_{mm'}
\end{align}
Using (\ref{68}) and in the $(\chi,\theta,\phi)$ parametrization of $S^3$,
 the action (\ref{69}) between the adjacent times $n$ and $n+1$ takes the form
\begin{align}\label{79}
S_E(n,n+1)=2\kappa\,\Big(\cos\!\chi_{n+1}\,\cos\!\chi_{n}+\bm{\hat{x}}_{n+1}\cdot\bm{\hat{x}}_n\, \sin\!\chi_{n+1}\, \sin\!\chi_{n}\,-1\Big)
\end{align}
in which 
\begin{align}\label{80}
\bm{\hat{x}}_{n+1}\cdot\bm{\hat{x}}_n=\cos \theta_{n+1} \cos\theta_n
+\cos(\phi_{n+1}-\phi_n) \sin\theta_{n+1}\sin \theta_n
\end{align}
Then the matrix elements of the transfer matrix is given by
\begin{align}\label{81}
\langle  \bm{x}_{n+1} |\widehat{V} |  \bm{x}_n\rangle= 
\sqrt{\frac{\kappa}{4\pi}} \exp\!\Big[ S_E(n,n+1)\Big]
\end{align}
Using the identity for real $w$
\begin{align}\label{82}
e^{w\, \bm{\hat{x}}\,\cdot\,\bm{\hat{x}}'}=  \sqrt{\frac{\pi}{2w}} ~
4\pi\sum_{\ell =0}^\infty \sum_{m=-\ell }^\ell  I_{\ell +1/2}(w) \,\,
Y_{\ell m}(\Omega)\,\,Y_{\ell m}^\star(\Omega')
\end{align}
in which the direction of two unit vectors $\bm{\hat{x}}$ and 
$\bm{\hat{x}}'$ are given by ordinary solid-angles $\Omega$ and $\Omega'$,
respectively, and $I_{\ell +1/2}$ as before is the modified Bessel function.
Taking $w=2\kappa\sin\!\chi_n\sin\!\chi_{n+1}$, by (\ref{82}) we have 
\begin{align}
\sqrt{\frac{\kappa}{4\pi}}  
\exp\!\Big[ S_E(n,n+1)\Big]=\sqrt{\frac{\kappa}{4\pi}} \,
e^{2\kappa(\cos\!\chi_n\cos\!\chi_{n+1}-1)} 
\sqrt{\frac{\pi}{4\kappa\sin\!\chi_n\sin\!\chi_{n+1}}} 
\cr
\label{83}
\times 4\pi\sum_{\ell =0}^\infty \sum_{m=-\ell }^\ell  I_{\ell +1/2}(2\kappa \sin\!\chi_n\sin\!\chi_{n+1}) \, Y_{\ell m}(\Omega_n)\,Y_{\ell m}^\star(\Omega_{n+1})
\end{align}
for which we also have an expansion based on the energy eigen-values 
and the eigen-functions as
\begin{align}
\sqrt{\frac{\kappa}{4\pi}} 
\exp\!\Big[ S_E(n,n+1)\Big]&= \sum_{s=0}^\infty \sum_{\ell =0}^s \sum_{m=-\ell }^\ell   
e^{-aE_{s\ell m}}\,  \mathbb{Y}_{s\ell m}(\chi_n,\Omega_n)
\,\mathbb{Y}_{s\ell m}^\star(\chi_{n+1},\Omega_{n+1})
\cr
\label{84}
&=\sum_{\ell =0}^\infty \sum_{m=-\ell }^\ell   \sum_{s=\ell }^\infty
e^{-aE_{s\ell m}}\,  \mathbb{Y}_{s\ell m}(\chi_n,\Omega_n)
\,\mathbb{Y}_{s\ell m}^\star(\chi_{n+1},\Omega_{n+1})
\end{align}
Using the orthonormality relation of $Y_{\ell m}$'s, by (\ref{83}) and (\ref{84}), 
after the changes $\chi_n\to\chi$ and $\chi_{n+1}\to\chi'$, explicit expression (76) gives us
\begin{align}
\sum_{s=\ell }^\infty
\frac{2^{2\ell +1}(s+1)\, (s-\ell )!\, \ell !^{\,2}}{\pi (s+\ell +1)!}
e^{-aE_{s\ell }} \sin^\ell  \!\chi \sin^\ell  \!\chi'\,
C^{(\ell +1)}_{s-\ell }\!(\cos\!\chi) C^{(\ell +1)}_{s-\ell }\!(\cos\!\chi') 
\cr  \label{85}
=\sqrt{\frac{\kappa}{4\pi}} 
\sqrt{\frac{\pi}{4\kappa\sin\!\chi\,\sin\!\chi'}}\, 4\pi\,
e^{2\kappa(\cos\!\chi \cos\!\chi'-1)} 
I_{\ell +1/2}(2\kappa \sin\!\chi \sin\!\chi')
\end{align}
\begin{figure}[t]
	\begin{center}
		\includegraphics[scale=0.9]{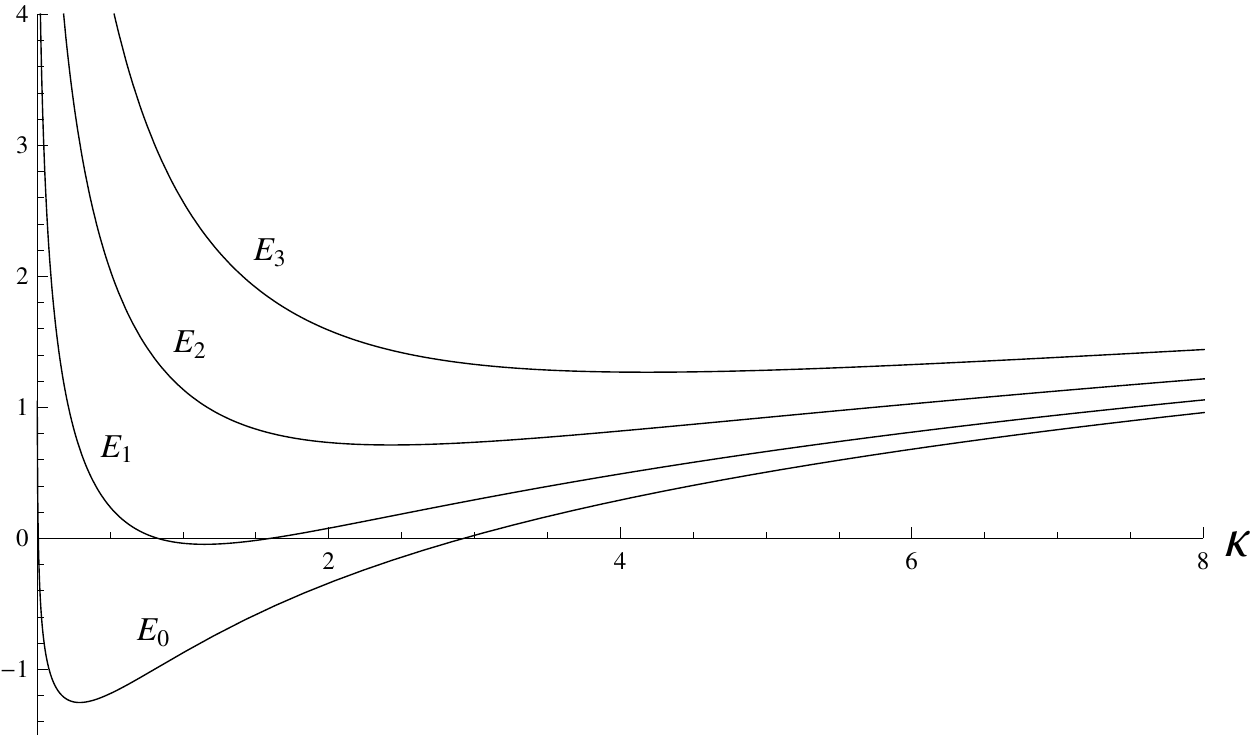}
	\end{center}
	\caption{The few lowest energies by (\ref{90}) for SU(2) group. }
\end{figure}
in which we have dropped the index $m$ in $E_{s\ell }$, as it is now an irrelevant one. 
This indicates that the energy eigen-values has 
at least $(2\ell +1)$-level degeneracy. Using the orthogonality
of the Gegenbauer polynomials
\begin{align}\label{86}
\int_0^\pi C^{(\ell+1)}_{s-\ell}\!(\cos\alpha)\, C^{(\ell+1)}_{s'-\ell}\!(\cos\alpha) 
\sin^{2\ell+2}\!\!\alpha 
\,d\alpha = \frac{\pi\, (s+\ell+1)!}{2^{2\ell+1}(s+1)(s-\ell)!\, \ell !^{\,2}}\, \delta_{ss'}
\end{align}
by multiplication the l.h.s. of (\ref{85}) by 
$\sin^{\ell+2}\!\chi' \, C^{(\ell+1)}_{s-\ell}(\cos\chi')$ and 
integration over $\chi'$  we have 
\begin{align}
e^{-aE_{s\ell }} \, \sin^\ell  \!\chi  \,
C^{(\ell +1)}_{s-\ell }\!(\cos\!\chi) 
= \int_0^\pi\!\! d\chi' 
\sqrt{\frac{\kappa}{4\pi}} 
\sqrt{\frac{\pi}{4\kappa\sin\!\chi\,\sin\!\chi'}}\, 4\pi\,
e^{2\kappa(\cos\!\chi\cos\!\chi'-1)} 
\cr   \label{87}
\times\sin^{\ell+2}\!\chi' \, C^{(\ell+1)}_{s-\ell}\!(\cos\chi')
I_{\ell +1/2}(2\kappa \sin\!\chi\sin\!\chi')
\end{align}
\begin{figure}[t]
	\begin{center}
		\includegraphics[scale=1.1]{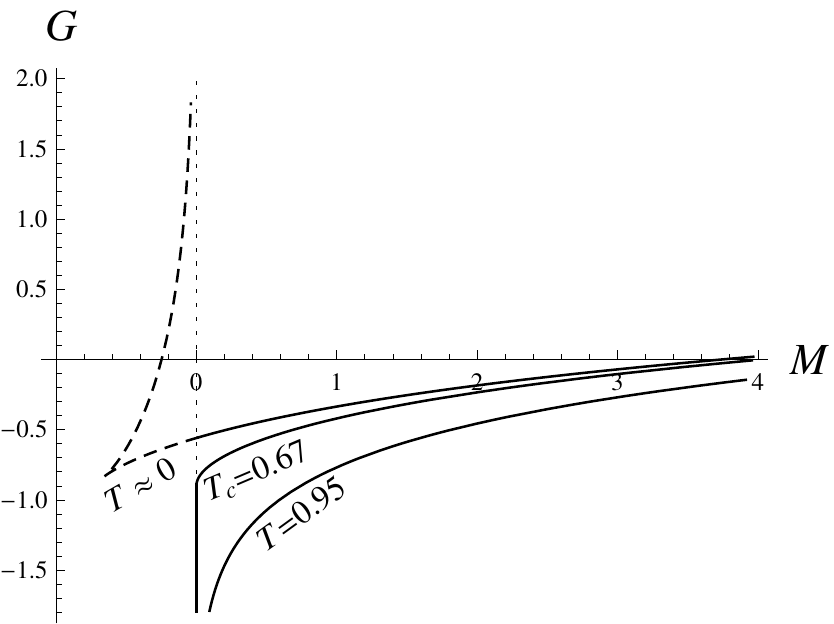}
	\end{center}
	\caption{The isothermal $G$-$M$ plots for SU(2) group. }
\end{figure}
Using the identity \cite{gegen}:
\begin{align}
\int_0^\pi e^{z \cos\chi \cos \chi'} \sin^{\ell+3/2}\!\chi'\,
C^{(\ell+1)}_{s-\ell}\!(\cos\!\chi') \,I_{\ell+1/2}(z\sin\!\chi \sin\!\chi')\,d\chi'
\cr    \label{88}
=\sqrt{\frac{2\pi}{z}}~ \sin^{\ell+1/2}\!\chi ~
C^{(\ell+1)}_{s-\ell}\!(\cos\!\chi) ~ I_{s+1}(z)
\end{align}
the integration in r.h.s. of (\ref{87}) can be calculated, setting $z=2\kappa$, 
leading to
\begin{align}\label{89}
e^{-a\,E_s} =\sqrt{\frac{\kappa}{4\pi}} \,
\frac{2\pi^2}{\kappa}\,
e^{-2\kappa} \, I_{s+1}(2\kappa)
\end{align}
in which we have dropped the index $l$ from $E_s$ as well, since there is no
$\ell$-dependence in spectrum. By (\ref{89}) we find the energy as below with
a $(s+1)^2$-level degeneracy: 
\begin{align}\label{90}
E_s=-\frac{1}{a}\ln \left[\sqrt{\frac{\kappa}{4\pi}} \,
\frac{2\pi^2}{\kappa}\,
e^{-2\kappa} \, I_{s+1}(2\kappa)
\right],~~~~s=0,1,2,\cdots
\end{align}
\noindent In Fig.~6 the four lowest energies are plotted, all 
having minimum. This is in contrast to
the cases with U(1) and $\mathbb{Z}_N$ groups
where only $E_0$ has a minimum.
The one-particle partition function is given by
\begin{align}
Z_1(\beta,\kappa)=\sum_{s=0}^\infty (s+1)^2~ e^{-\beta\,E_s}
\end{align}
Again the equation-of-state as well as the Gibbs free energy can be obtained by 
(\ref{43}) and (\ref{44}). The $G$-$M$ plots given in Fig.~7 develop cusp, indicating that 
the system exhibits a first-order phase transition below the critical temperature $T_c=0.67$.
At $T\approx 0$ the critical $\kappa_c=0.29$ is obtained above which $M$ gets non-zero values. 

\section{Conclusion and Discussion}
The general theoretical ground as well as specific examples are presented 
for models based on compact angle coordinates. 
The present construction might be considered as a continuation of the 
theme by which the gauge fields are treated as compact angle variables \cite{polya1,polya2,thooft1,lattice}. In the present formulation the action 
depends on the group elements rather than the
algebra elements, leading to the invariance under the total shifts inside 
the compact domain. It is observed that a discrete-time 
formulation of the theory is the 
natural way to obtain the desired dependence on group elements. 
The way toward the present formulation is in some sense in reverse 
direction of what has happened in lattice formulation of gauge theories.
In particular, in formulation of gauge theories on lattice the invariance 
under the gauge transformations requires that the gauge fields are introduced 
to the theory via the group elements. Here the demand of invariance under 
the total shifts of compact coordinates requires to treat time parameter 
as a discrete one. The worldline action by the formulation
resembles the spin chain Hamiltonian of magnetic systems, with coordinates appearing 
as the spin degrees of freedom.  

The quantization of the model is formulated based on the transfer-matrix method 
\cite{lattice,wipf}. The pre-factor in the definition of 
the elements of transfer-matrix, in contrast to the case with infinite extent coordinates, can not be absorbed by a change of the path-integral 
integration variables. 
This particularly causes that the energy eigen-values develop minima as
functions of the defining parameter of the theory.

As examples for the formulation, the models based on the U(1), 
$\mathbb{Z}_N$ and SU(2) groups are explicitly constructed.
In all of the models based on the three groups the exact energy 
eigen-values are obtained. As the consequence of the minima in the spectrum
all the models exhibit the first-order transition between coexistent phases.

As mentioned earlier, by setting $\kappa=1/g^2$ the model by U(1) group is in fact the result of the dimensional reduction of pure U(1) lattice gauge theory. As one possible application of the present construction here we mention the attempt in \cite{0bfath} to fit the model by U(1) group to the expectations from the monopole dynamics. In particular, the phase transition for the particles with mass $m_0\propto 1/g^2$ by (\ref{25}) may provide a better understanding of the role of monopoles in confinement mechanism based on the dual Meissner effect in superconductors \cite{nambu,mandelstam,thooft2}. Based on the proposed mechanism, at strong coupling limit, at which the monopoles have tiny masses, the motion of monopoles around the electric fluxes prevents the fluxes to spread, leading to the confinement of the electric charges.  Instead at small coupling limit, where the monopoles are highly massive, the electric fluxes originated from source charges are likely to spread over space, leading to the Coulomb's law. It is expected that there is a critical coupling $g_c$ at which the transition from confined phase to the Coulomb phase occurs.

According to the model with group U(1), the two regimes with weak and strong couplings constants are related by a first-order phase transition. The behavior of the system at low temperatures, where the main contribution to the partition function is from the ground-state, is of particular interest. In the limit $T\to 0$, due to the Maxwell construction, we have $M=0$ for $g<g_c=1.125$; Fig.~8. 
\begin{figure}[t]
	\begin{center}
		\includegraphics[scale=1.1]{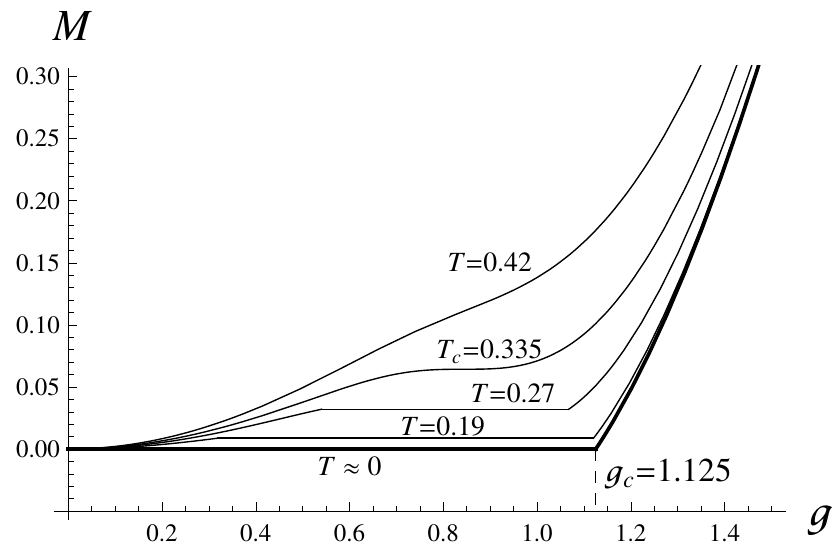}
	\end{center}
	\caption{The isothermal $M$-$g$ plots by the model based on the U(1) group.}
\end{figure}
So as the consequence of discontinuous nature of the first order transition, at low temperatures and below $g_c$ we have $M\!\propto\! \langle v^2\rangle\approx 0$. This behavior is to be compared with (\ref{46}) for ordinary particles, by which there is an asymptote reduce of $M$ by increasing the mass at constant $T$. According to the present model, at low temperatures and below $g_c$, the particles with mass $m_0\propto g^{-2}$ are hardly moving ($\langle v^2 \rangle \approx 0$), leading to an exact Coulomb phase. On the other hand, exhibiting a high-slope increase of $\langle v^2 \rangle$ at $g_c$, the confined phase finds an instant govern once $g$ exceeds $g_c$ at low temperatures. This picture and specially the value of critical coupling constant are in agreement with theoretical and numerical studies \cite{0bfath}. 

\vspace{4mm}
\textbf{Acknowledgement}: The author is grateful to M.~Khorrami
for helpful discussions on the role of the imaginary time in formulation of 
gauge theories on lattice. 
This work is supported by the Research Council of the Alzahra University.


\end{document}